# Digital PCR provides sensitive and absolute calibration for high throughput sequencing

Richard A. White III#, Paul Blainey#, H. Christina Fan and Stephen R. Quake*

*Department of Bioengineering, Stanford University and Howard Hughes Medical Institute, Stanford, California 94305*

**Abstract** Several of the next generation sequencers are limited in their sample preparation process by the need to make an absolute measurement of the number of template molecules in the library to be sequenced. As currently practiced, the practical effects of this requirement compromise sequencing performance, both by requiring large amounts of sample DNA and by requiring extra sequencing runs to be performed. We used digital PCR to quantitate sequencing libraries, and demonstrated its sensitivity and robustness by preparing and sequencing libraries from subnanogram amounts of bacterial and human DNA on the 454 and Solexa sequencing platforms. This assay allows absolute quantitation and eliminates uncertainties associated with the construction and application of standard curves. The digital PCR platform consumes subfemptogram amounts of the sequencing library and gives highly accurate results, allowing the optimal DNA concentration to be used in setting up sequencing runs without costly and time-consuming titration techniques. This approach also reduces the input sample requirement more than 1000-fold: from micrograms of DNA to less than a nanogram.

*To whom correspondence should be addressed. E-mail: quake@stanford.edu. Phone (650) 724-7890 Fax (650) 736-1961
# These authors contributed equally to this work

**Introduction** A new generation of sequencing technologies based on "sequencing by synthesis" are revolutionizing biology, biotechnology, and medicine. A key advance facilitating higher throughput and lower costs for several of these platforms was migration from clone-based sample preparation commonly used in Sanger sequencing to massively parallel clonal PCR amplification of sample molecules on beads (Roche 454 and ABI Solid) (Margulies *et al*, 2005) or on a surface (Solexa) (Bing *et al*, 1996). The parallel amplification steps are relatively efficient, with sequence data obtained from a significant fraction of sequencing library molecules used in the amplification. This, when coupled with a good loading efficiency onto the instrument results in on the order of one million library molecules (typically less than a picogram of library DNA) being required to carry out a full sequence run. However, the manufacturers require one to ten trillion (typically 1 – 5 micrograms) DNA fragments as input for library preparation. This is primarily because quantitation of the library DNA according to the manufacturers' protocols consumes more than a billion molecules, and secondarily because of the limited efficiency of the library preparation methods, which have typical conversion efficiencies of 0.001% - 0.1%.

The requirement for micrograms of input DNA limits the pool of samples that can be analyzed with next generation sequencing technologies, since for many applications microgram quantities of sample are not available. In some cases it is possible to use amplification such as PCR or MDA, but amplifiers have bias and introduce distortion. We developed a method for highly accurate absolute quantitation of sequencing libraries that consumes subfemptogram amounts of library material based on digital PCR (Vogelstein 1999). Eliminating the large quantity requirement for traditional quantitation has the direct effect of reducing the sample input requirement from micrograms to nanograms or less, opening the way for analysis of minute and/or precious samples onto the next-generation sequencing platforms without the distorting effects of pre-amplification (Mackelprang *et al.* 2008). It also eliminates the need for expensive titration sequencing runs, which most manufacturers currently recommend.

**Results:** TaqMan detection chemistry has the advantage of yielding a fluorescence signal proportional to the number of molecules that have been amplified, not by the total mass of dsDNA in the sample (Heid *et al.* 1996). This method works by the addition of a double-labeled oligonucleotide probe in a PCR reaction powered by a polymerase with 5' to 3' exonuclease activity.

The probe must be complimentary to one of the two product strands such that the extending polymerase will encounter it and separate the two labels by exonuclease activity, activating the probe's fluorescence. Conventional TaqMan detection chemistry requires that the probe is complementary to the region within the amplified portion of the template between the two amplification primers. This strategy is not possible for the sequencing libraries, which have inserts of unknown or random sequence between short adaptor sequences. To overcome the challenge of probe design for templates of random sequence, we adopted the universal template (UT) approach where a probe-binding sequence is appended to one of the PCR primers (Zhang *et al.* 2003). To decrease reaction times, we replaced the published 20 bp UT probe-binding region with an 8 bp sequence target for a probe containing a locked nucleic acid nucleotide as applied in Roche's UPL (Universal Probe Library) probes. The shorter amplicon-probe interaction length allows the reduction of PCR run times from 2.5 hours to less than 50 minutes. In practice, we often use the UT-qPCR assay in the real-time mode (with a calibration standard) to range the library concentration so that an appropriate dilution can be made for absolute quantitation by UT-digital PCR.

Digital PCR gives an absolute, calibration-free measurement of the concentration of amplifiable library molecules, with a lower coefficient of variation than a real-time PCR measurement with an ideally prepared standard curve (Figure 1A). We have sequenced more than thirty 454 libraries without a single titration run over a five month period using the digital PCR quantitation method. (Figure 2; selected libraries are described in Tables 1 and 2) These libraries, despite dramatic differences in source, type, molecular weight, and quality, all gave ideal emPCR and enrichment/sequencing results with a very narrow range of DNA to bead ratios, typically about 0.1 DNA per bead (Figure 2).

To demonstrate the utility of digital PCR in preparing DNA libraries from small amounts of starting material, twelve libraries were created from starting amounts of *E. coli* DNA ranging from 35 ng to as low as 500 pg. Six of the libraries were constructed with *E. coli* genomic DNA (with prior dilution before construction of library) with the standard 454 shotgun protocol with Roche's molecular barcodes, "MIDs" (or Multiplex IDentifiers). Six more DNA libaries were prepared from the same quantities of an *E. coli* amplification product (of 466 bp). The DNA libraries were quantified via UT dPCR as described.

The results from Figure 1B show that we can obtain enough library DNA from 500 pg of genomic (shotgun) or amplicon DNA to create more than 100,000 enriched beads for sequencing. All twelve trace libraries were sequenced in a full run of our GS FLX 454 DNA pyrosequencer. In total, 18 million raw bases were sequenced from the trace shotgun libraries and 38 million raw bases were sequenced from the amplicon libraries. 69.16% of the shotgun reads and 99.17% of the amplicon reads mapped back to *E. coli*. Specifically, in the case of the library made from 500 pg of *E. coli* 16S amplicon, half of the resulting library was used for sequencing. 14.0 million raw bases were obtained in 55,206 reads with 99.02% of the reads mapping back to the template, indicating that almost 30 Mbp can be obtained from a library of 131,000 molecules prepared from 500 pg input material. Similarly, half of the 1 ng *E coli* amplicon library gave 10.9 million raw bases in 43,217 reads with 99.17% mapping. The 500 pg *E coli* shotgun library gave 5.7 million raw bases in 26,812 reads (69.9% mapping), while the 1 ng *E coli* shotgun library gave 6.0 million raw bases in 28,730 reads (69.9% mapping).

In an earlier shotgun sequencing run we used 2,400,000 sstDNA fragments (or 0.71 pg amplifiable DNA) from an *Acetonemia longum* shotgun library DNA (prepared according to the standard library preparation method from 723 ng of genomic DNA). From these molecules, accurately and reproducibly quantitated by digital PCR, 74% of the beads loaded gave useful 454 sequence data (4.13% 'mixed' reads and 4.28% 'dot' reads) to yield 67 Mbp in 278,181 reads on one large PTP region (one-half of the 454 *FLX* sequencing run). Together with 38 Mbp from another run, 105.6 Mbp of very high quality *Ace* shotgun data were obtained without any titration techniques, 104.3 Mbp of which assembled under Newbler to give coverage of the ~5 Mbp *Acetonemia longum* genome with N50 contig size greater than 50,000 bp. Here we have demonstrated that significant quantities of DNA pyrosequencing data can be obtained from subnanogram DNA samples without titration runs.

A similar UT-dPCR assay was designed to quantify Solexa sequencing libraries. Solexa libraries were prepared from human plasma DNA or whole blood genomic DNA using starting amounts of DNA between 2 and 6 ng. The concentrations of library molecules were determined by UT-dPCR, and diluted to 4 pM for loading onto the sequencer. We achieved consistent cluster density between ~110,000 to 150,000 clusters per tile on the Genome Analyzer II, a range that is deemed optimal by the manufacturer. The total number of reads yielded was 11 to 15 million per lane (Table 2). The libraries were also quantitated on the Agilent Bioanalyzer and NanoDrop spectrophotometers. Had we determined the dilutions based on these standard techniques, we would have obtained cluster densities too high and too low by factors of two, respectively.

**Discussion** The standard workflow for the next-generation instruments entails library creation, (requiring a bulk PCR step on Solexa), massively parallel PCR amplification of library molecules, and sequencing. Library creation starts with conversion of the sample to appropriately sized fragments, ligation of adaptor sequences onto the ends of the sample molecules, and selection for molecules properly appended with adaptors. The presence of the adaptor sequences on the ends of the library molecules enables amplification of random-sequence inserts by PCR. The number of library DNA molecules in the massively parallel PCR step is critical: it must be low enough that the chance of two associating with the same bead (454) or the same surface patch (Solexa) is low, but, there must be enough library DNA present such that the yield of amplified sequences is sufficient to realize a high sequencing throughput. The standard workflow calls for measuring the mass of library DNA using the Agilent Bioanalyzer capillary gel electrophoresis (GE) instrument (454) or the nanodrop spectrophotometer (Solexa), and then converting the mass to a molecule count using knowledge of the length distribution.

Quantification of the library by mass presents three major stumbling blocks that render the quantification inaccurate to the degree where the sequencing results can be adversely affected. First, mass-based quantitation also requires an accurate estimate of the length of the molecules to determine the molar concentration of DNA fragments. Second, degraded and damaged molecules that cannot be amplified in the massively parallel amplification step are counted. And third, methods of measuring DNA mass lack sensitivity, and are imprecise in concentration measurements near the limit of detection.

When the library concentration is underestimated, the possibility of molecular crosstalk arises due to the clonality of beads (454) or clusters (Solexa) being compromised, which reduces the fraction of useful reads. When the library concentration is overestimated, the number of beads recovered (454) or number of clusters generated (Solexa) is reduced, in which case the full capacity of the sequencers cannot be utilized. Before carrying out a bulk sequencing run with a new library, Roche and Illumina recommend carrying out a four-point titration run on their sequencers in order to empirically determine the optimal volume of DNA for the massively parallel PCR. In addition,

Illumina recommends that the user check the library quality with traditional Sanger Sequencing before its application for high-throughput sequencing. Digital PCR method eliminates all three of these problems and the requirement for titration.

Quantitative Real-time PCR, and especially digital PCR, are ideal candidate techniques for this application because of their exquisite sensitivity. Some detection chemistries for real-time PCR, such as TaqMan, have the property of counting molecules rather than measuring DNA mass, although the measurements are relative and the methods by which standards are established often tie the real-time PCR quantitation back to sample mass. Digital PCR is a technique where a limiting dilution of the sample is made across a large number of separate PCR reactions such that most of the reactions have no template molecules and give a negative amplification result. In counting the number of positive PCR reactions at the reaction endpoint, one is counting the individual template molecules present in the original sample one-by-one. A major advantage of digital PCR is that the quantitation is independent of variations in the amplification efficiency – successful amplifications are counted as one molecule, independent of the actual amount of product. PCR-based techniques have the additional advantage of only counting molecules that can be amplified, *e.g.* that are relevant to the massively parallel PCR step in the sequencing workflow. We use Fluidigm's Biomark platform for digital PCR, which performs 9,180 PCR reactions per chip with automated partitioning of nanoliter PCR reactions on 12 independent input samples.

Recently, Meyer *et al.* developed a SYBR Green real-time PCR assay that allows the user to estimate the number of amplifiable molecules in sequencing trace samples. This was the first report of PCR-based quantitation of sequencing libraries, and extended the sensitivity of library quantitation significantly - although to an unknown extent, since the source material used to make the trace libraries was not quantitated. However, the SYBR Green assay presents principle disadvantages: 1) SYBR Green I dye is an intercalating flurochrome that gives signal in proportion to DNA mass, not molecule number, 2) SYBR Green assays rely on an external standard that limits the absolute accuracy over time and is not universal to all sample types, and 3) intercalating fluorochomes give signal from nonspecific PCR reaction products.

In a real-time assay, the standard must have the same amplification efficiency and molecular weight distribution as the unknown library sample. This means the user must have on hand a bulk sequencing library very similar to the trace library being made and that the molecular weight distributions of both the standard and the new library be known—often impractical requirements for a trace sample library. Furthermore, this standard library must be of extremely high quality if mass-based quantitation is to be used to calibrate the assay for amplifiable molecules, which makes assessment of the concentration of amplifiable molecules in a degraded sample extremely difficult.

Lastly, sequence-nonspecific detection chemistries like SYBR Green give signal from all dsDNA products generated, including primer dimers and nonspecific amplification products, which may be an issue in complex samples. In particular, side products can compete with specific amplification from low numbers (<1000) of template molecules, limiting the accuracy of SYBR Green quantitation for dilute samples (Simpson 2000). Although the presence of these side products can often be discerned by analysis of the product melting curve, opportunities to optimize the primers are limited due to the short length of the adaptor sequences and the specific nucleotide sequences required for compatibility with proprietary sequencing reagents. Sensitivity to side products gives SYBR Green a tendency toward overestimation of the sample quantity.

**Conclusion** This work presents an assay that circumvents these limitations using TaqMan detection chemistry and digital PCR. When combined with digital PCR, dependence on a standard sample is eliminated, and the results are sufficiently accurate to allow the elimination of titration techniques, even for samples of low quantity and low quality. The extreme sensitivity of real-time and digital PCR eliminate quantitation as the material-limiting step in the sequencing workflow, bringing greater focus to library preparation procedures as the most limiting step in sequencing trace samples. It is natural to expect that library preparation procedures developed with the capacity to handle up to five micrograms of input are far from optimal with respect to minimizing loss from nanogram or picogram samples. Library preparation procedures optimized for trace samples with reduced reaction volumes and media quantities, possibly formatted in a microfluidic chip, have the potential to dramatically improve the recovery of library molecules, allowing preparation of sequencing libraries from quantities of sample comparable to that actually required for the sequencing run, *e.g.* close to or less than one picogram.

Digital PCR quantitation is sufficiently accurate in counting amplifiable library molecules to justify elimination of titration techniques as well as the associated cost and time involved. The method is also hundreds of millions of times more sensitive than traditional means of library quantitation, and allows the sequencing of libraries prepared from tens to hundreds of picograms of starting material, rather than the micrograms of DNA required by the manufacturers' protocols. The reduced sample requirement enables the application of next-generation sequencing technologies to minute and precious samples without the need for additional amplification steps.

**Methods:**

**Sample generation.**
DNA was extracted/isolated for mid-log phase K12 over night cultures using Qiagen's DNeasy Tissue & Blood kit then further purified using Qiagen's QIAquick PCR purification kit following the manufacturer's protocol. *E.coli* amplicons were generated from 16s rRNA PCR following standard protocols to generate a uniform 466 bp fragment. DNA extracted from human plasma or whole blood using Qiagen's DNA Blood Mini Kit or Machinerey-Nagel's NucleoSpin Plasma Kit according to manufacturers' protocols.

**Sequencing library preparation.**
454 shotgun libraries: Libraries were generated according to the manufacturer's protocol with a few adjustments: trace *E.coli* amplicons and human sample pX were not nebulized; 0.01% Tween-20 was added to the elution buffer for each mini-elute column purification step; libraries were eluted using 1xTE containing 0.05% Tween-20 at a volume of 30 µl. Single stranded libraries were aliqouted for storage.

Solexa libraries: Libraries were generated following standard genomic DNA protocol with small adjustments. No nebulization was performed on plasma DNA samples since they were fragmented in nature (average ~170 bp); the whole blood genomic DNA sample was sonicated to produce fragments between 100 and 400 bp; all ligated products were used for 18-cycle PCR enrichment; no gel extraction was performed and no Sanger sequencing was used to confirm fragments of correct sequence.

**Standard creation for UT-qPCR for the Statagene Mx3005.**
After sequencing library preparation, UT-qPCR was used to range the concentration for UT-dPCR. For testing purposes and to gauge the correct dilution, a standard library was created, quantitated on UT-dPCR and serially diluted for UT-qPCR calibration. In order to ensure uniform amplification among various libraries, the fragment length distribution of the standard matched the that of the library to be quantitated. To maintain the standard over time, the library was cloned into pCR2.1 (Invitrogen) and then transformed into DH5α cells. Plasmids containing library standard were harvested from mid-log phase DH5α cells and then further isolated using Qiagen's QIAprep Spin Miniprep kit. The resulting plasmids were digested using EcoRI, then gel purified and cleaned up using Qiagen's QIAquick PCR purification kit. Calibration of the UT-dPCR of the standard was conducted on a regular basis.

**UT-qPCR quantitation on the Statagene's Mx3005.**
Validated standards were diluted in ten-fold increments through the range $10^{15}$-$10^{3}$ molecules/µl. Standards were assayed in triplicate in order to obtain standard deviation/relative coefficient of variation. Each library was diluted ten-fold, and assayed with twelve replicates in order to obtain standard deviation/relative coefficient of variation. The thermal cycling parameters are listed below.

**UT-dPCR quantitation on Fluidigm's BioMark System.**
454 libraries: UT-qPCR was first performed on aliquoted libraries in order to estimate the dilution factor for UT-dPCR. The libraries were diluted to roughly 100-360 molecules per µl. PCR reaction mix containing the diluted template was loaded onto Fluidigm's 12.765 Digital Array microfluidic chip. The microfluidic chip has 12 panels and each panel contains 765 chambers. The concentration of diluted template that yielded 150-360 amplified molecules per panel was chosen for technical replication. Six replicate panels on the digital chip were assayed in order to obtain absolute quantitation of the initial concentration of library. The diluted samples having typical relative coefficients of variation (between replicates) within 9-12% (or lower) were used for emPCR.

Solexa libraries: quantitative qPCR using human specific primers were first performed to estimate the dilution factor required for carrying out UT-dPCR. The final dilution yielded ~150-360 amplified molecules per panel.

All libraries: Reagents used for all UT-qPCR/UT-dPCR assays consisted of Universal Taqman Probe Master Mix (Roche) at 1x final concentration, 200 nM forward primer, 200 nM UT probe-binding primer, 400 nM reverse primer and 350 nM UPL (Universal Probe Library) #149 (Roche). The primer and probe sequences and the thermal cycling parameters are presented below.

**emPCR/Bridge PCR & Sequencing.**
454 sequencing: Sequencing was performed according to manufacturer's protocol. No titration or traditional sequencing was used. The DNA:bead ratios of 0.085 – 0.300 (based on UT-dPCR quantitation) were used. These ratios yielded the desired 10-15% bead recovery in enrichment and the lowest mixed sequence fraction. Mixed reads in 454 sequencing are defined as four consecutive positive nucleotide flows for a given read.

Solexa sequencing: Sequencing libraries were first diluted to 10 nM according to the concentration determined by digital PCR. The average dilution factor was 10 - 20. Diluted libraries were denatured with 2 N NaOH and then diluted to a final concentration of 4 pM. The templates were loaded onto flow cells. Cluster generation was performed according to the manufacturer's instructions. Sequencing was carried out on the Genome Analyzer II. No titration run was performed.

Thermocycling Parameters for UT-qPCR/UT-dPCR

| | Standard Adapters 454 UT-dPCR & UT-qPCR | MIDs/Paired-end UT-qPCR | MIDs/Paired-end UT-dPCR | Solexa UT-dPCR |
|---|---|---|---|---|
| Hot Start | 95C, 3mins | 95C, 3 mins | 95C, 3 mins | 95C, 10 mins |
| Denaturation | 94C, 30 secs | 95C, 3 secs | 95C, 15 secs | 95C, 15 secs |
| Annealing | 60C, 30 secs | 65C, 30 secs | 65C, 30 secs | 60C, 1 min |
| Extension | 72C, 45 secs | - | - | - |
| Cycle | 40 | 40 | 40 | 40 |

Primer/probe list for UT-qPCR/UT-dPCR

| | |
|---|---|
| | **Primers for Standard 454 libraries:** |
| **Forward:** | **5'-CCATCTCATCCCTGCGTGTC-3'** |
| **Reverse:** | **5'-CCTATCCCCTGTGTGCCTTG-3'** |
| **UTBP-1:** | **5'-GGCGGCGACCATCTCATCCCTGCGTGTC-3'** |
| | **Primers for 454 MID/Paired end libraries:** |
| **Forward:** | **5'-GCCTCCCTCGCGCCATCAG-3'** |
| **Reverse:** | **5'-GCCTTGCCAGCCCGCTCAG-3'** |
| **UTBP-2:** | **5'-GGCGGCGAGCCTCCCTCGCGCCATCAG-3'** |
| | **Primers for Solexa libraries:** |
| **Forward:** | **5'-ACACTCTTTCCCTACACGA-3'** |
| **Reverse:** | **5'-CAAGCAGAAGACGGCATA-3'** |
| **UTBP-3:** | **5'-GGCGGCGAACACTCTTTCCCTACACGA-3'** |
| **UPL#149** | **Universal probe sequence:** **5'-CCGCCGCT-3'** |

**Acknowledgments**: We thank Angela Wu, Jared Leadbetter, Liz Otteson, Matthias Meyer, Baback Gharizadeh, Farbod Babrzaeh, and Roxana Julili for assistance, helpful discussions and sharing 454 library samples. We also thank Joseph Derisi, Clement Chu, and Nick Ingolia for their help in carrying out sequencing experiments on the Solexa Genome Analyzer. This work was supported by Pioneer funding from the NIH to SRQ.

## Figure Legends

**Figure 1:** Reproducibility of UT-PCR assays. **A**. Twelve 454 Libraries were assayed with six to eight replicates by both UT-dPCR and UT-qPCR. UT-qPCR was calibrated using a library quantitated by digital PCR. The CV for dPCR is significantly lower than that for qPCR. **B**. Accurate digital PCR quantitation of 454 libraries from trace amounts (500 pg to 35 ng) of input *E. coli* genomic or amplicon DNA. Useful numbers of library molecules are recovered in all cases.

**Figure 2**: 454 enrichment and sequencing results. **A**: Histogram of bead enrichment fractions obtained in 454 sample preparation when digital PCR is used for quantitaiton. The manufacturer's recommended range is 10% to 15%, and our results range between 3% and 20%. **B**: Histogram of mixed fraction from 454 sequencing runs using samples calibrated by digital PCR. The manufacturer specifies the acceptable range to be 20% to 30% and our results range between 4% and 31%.

**Figure 3** Detection of three trace CHIP 454 sst DNA libraries prepared from 40 -60 ng input mouse chromatin by digital PCR. **A**, No signal from libraries on the NanoDrop spectrophotometer or the Agilent Bioanalyzer capillary electrophoesis unit. The two signals appearing in the electropherograms are molecular weight markers of 15 bp and 1500 bp. **B,** Detection of library molecules by digital PCR. False-color image of 12.765 digital array at assay endpoint. Each grid point corresponds to a nanoliter-scale PCR reaction, with yellow squares revealing amplification due to the presence of at least one sequencing library template molecule. The panels show dilution series (indicated) of samples analyzed in part A, allowing accurate absolute quantification of the sample by UT digital PCR.

**Figure 1**

**A:**

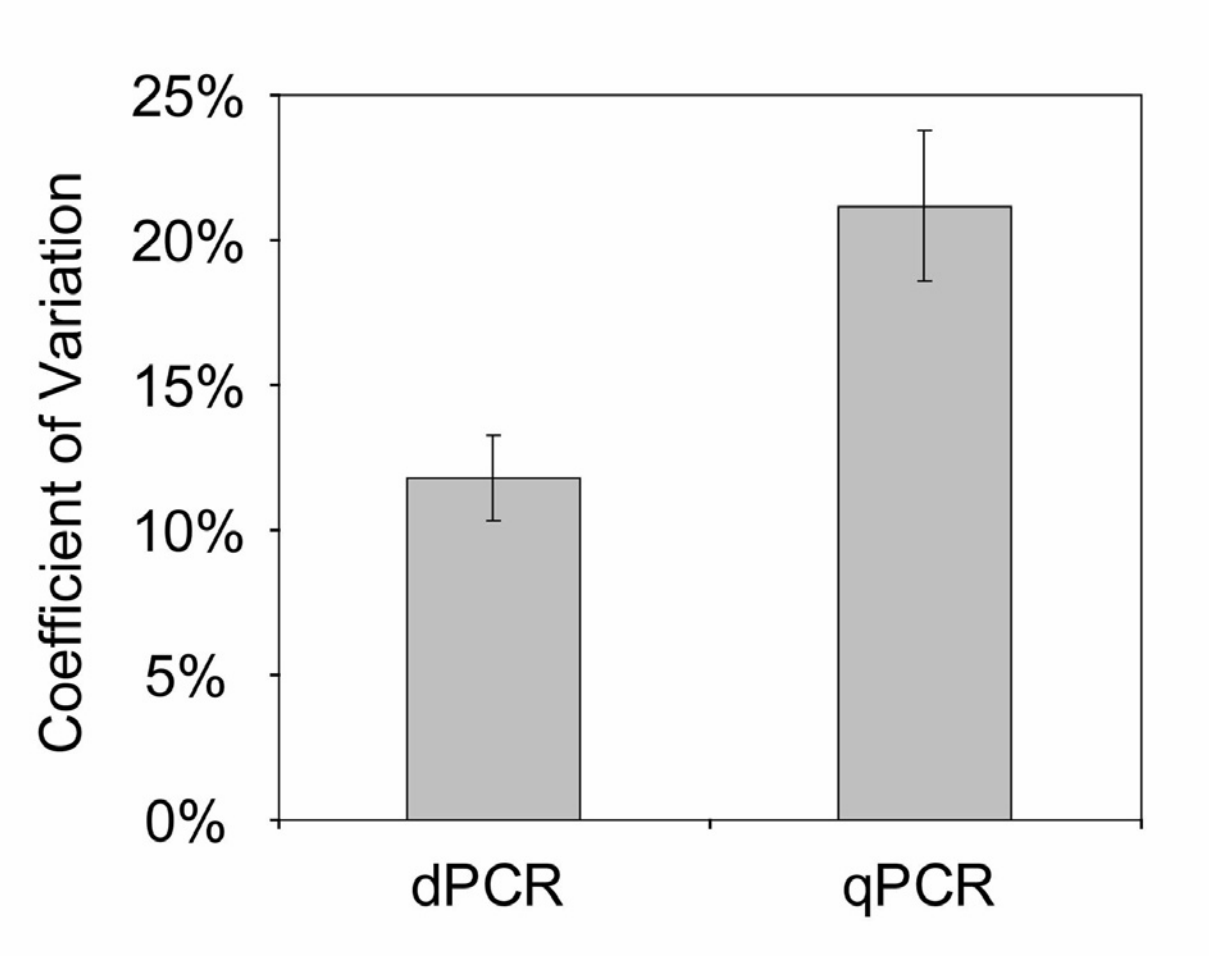

**B:**

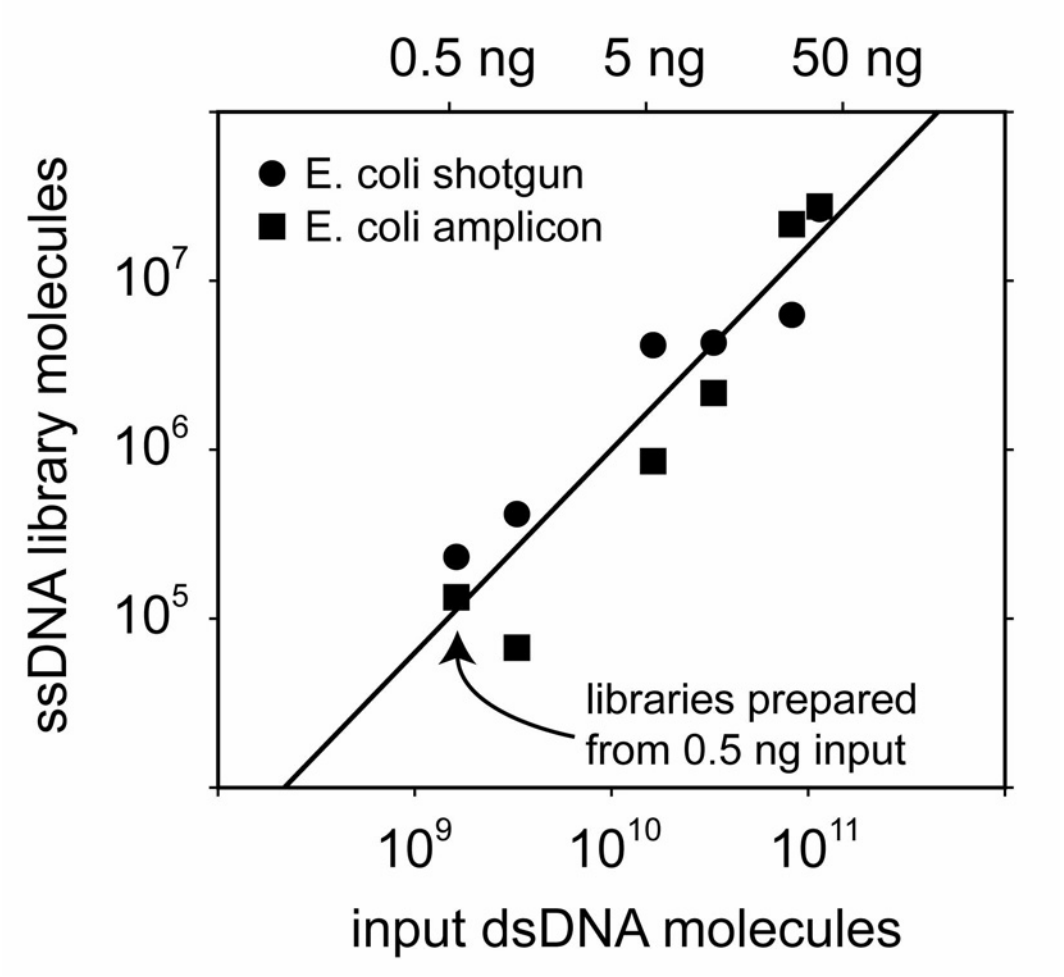

**Figure 2**

**A:**

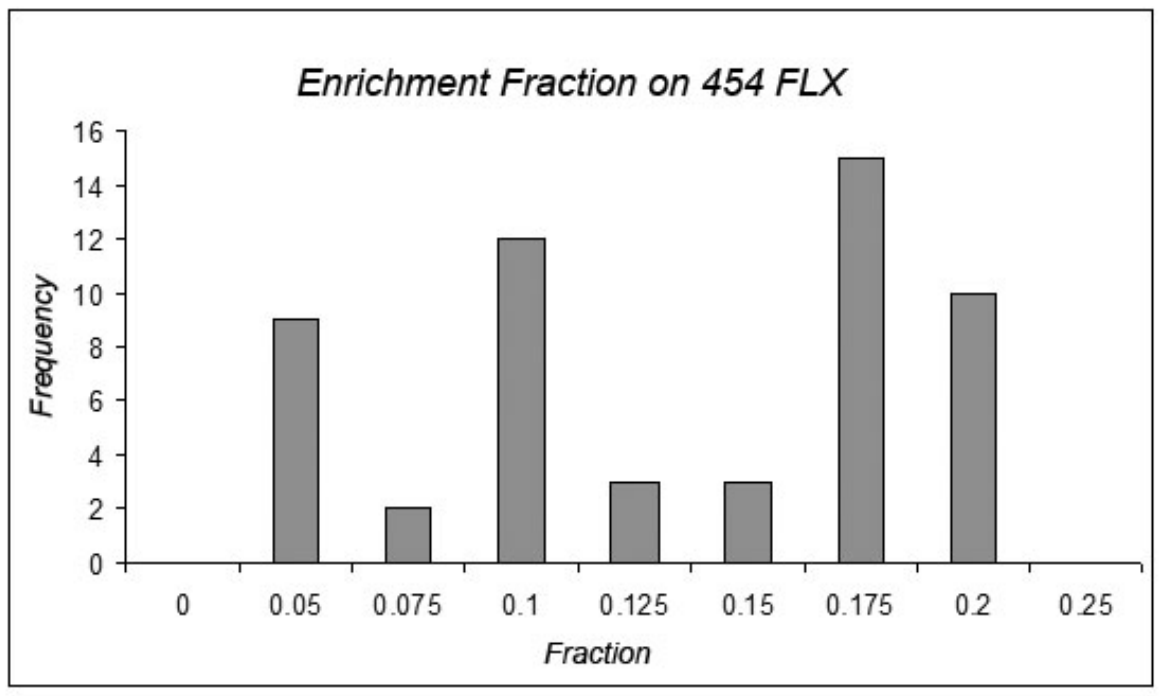

**B:**

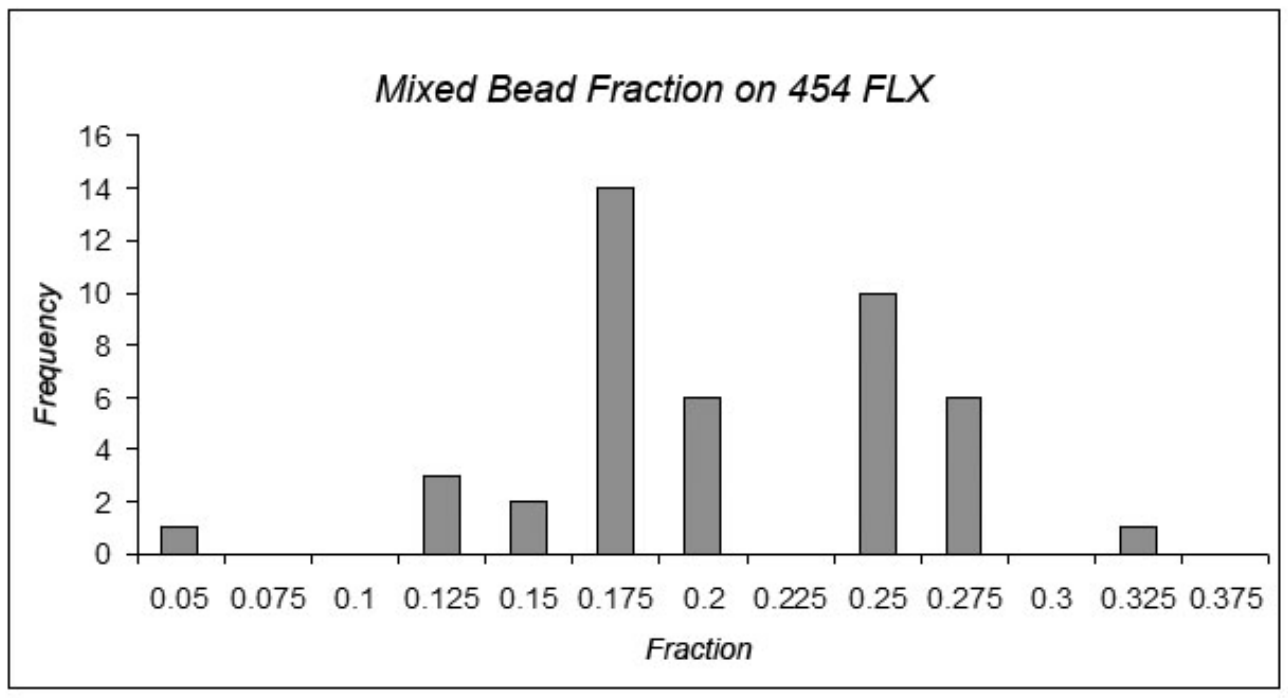

**Figure 3**

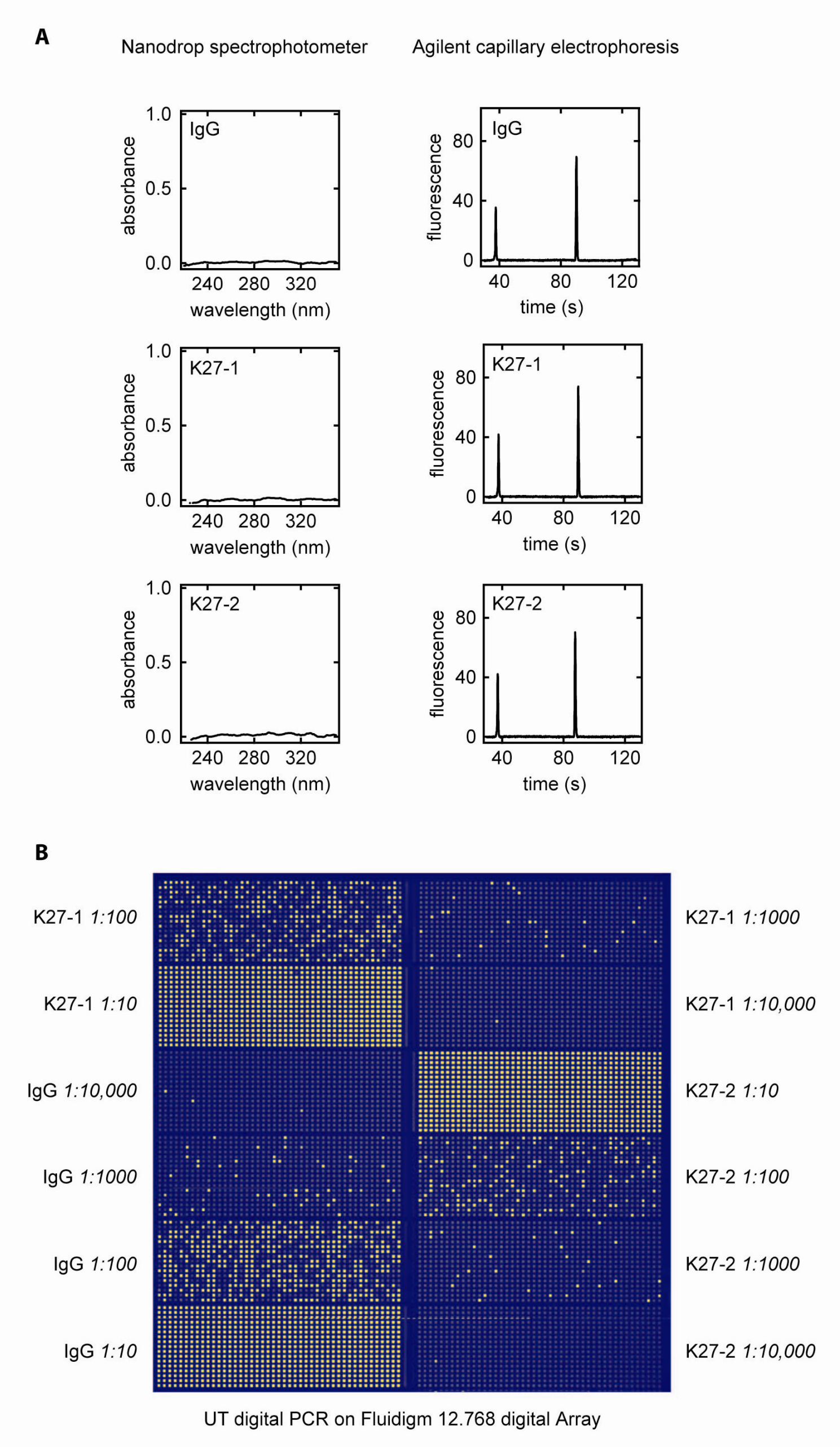
A
Nanodrop spectrophotometer
Agilent capillary electrophoresis
absorbance
wavelength (nm)
fluorescence
time (s)
IgG
K27-1
K27-2
B
K27-1 1:100
K27-1 1:10
IgG 1:10,000
IgG 1:1000
IgG 1:100
IgG 1:10
K27-1 1:1000
K27-1 1:10,000
K27-2 1:10
K27-2 1:100
K27-2 1:1000
K27-2 1:10,000
UT digital PCR on Fluidigm 12.768 digital Array

## Tables

### Table 1: Trace Microbial/Human 454 library Construction both Amplicon/Shotgun formats

| *Sample ID* | *input (ng)* | *Mean Library fragment size (Bp)* | *Input (total molecules)* | *ssDNA library (total molecules by UT-dPCR)* | dPCR replicate CV | *Library Prep Recovery %* | *Type* | *Organism* |
|---|---|---|---|---|---|---|---|---|
| ***TS-1*** | 35 ng | 500 | $1.17 \times 10^{11}$ | $2.61 \times 10^{7}$ | 7.07% | 0.022% | Shotgun | *E.coli* |
| ***TS-2*** | 25 ng | 500 | $8.37 \times 10^{10}$ | $6.15 \times 10^{6}$ | 8.24% | 0.007% | Shotgun | *E.coli* |
| ***TS-3*** | 10 ng | 500 | $3.35 \times 10^{10}$ | $4.20 \times 10^{6}$ | 7.71% | 0.013% | Shotgun | *E.coli* |
| ***TS-4*** | 5 ng | 500 | $1.67 \times 10^{10}$ | $4.08 \times 10^{6}$ | 5.65% | 0.024% | Shotgun | *E.coli* |
| ***TS-5*** | 1 ng | 500 | $3.35 \times 10^{9}$ | $4.08 \times 10^{5}$ | 16.84% | 0.012% | Shotgun | *E.coli* |
| ***TS-6*** | 500 pg | 500 | $1.67 \times 10^{9}$ | $2.31 \times 10^{5}$ | 5.58% | 0.014% | Shotgun | *E.coli* |
| ***TS-7*** | 35 ng | 466 | $1.17 \times 10^{11}$ | $2.67 \times 10^{7}$ | 8.55% | 0.023% | Amplicon | *E.coli* |
| ***TS-8*** | 25 ng | 466 | $8.37 \times 10^{10}$ | $2.11 \times 10^{7}$ | 8.50% | 0.025% | Amplicon | *E.coli* |
| ***TS-9*** | 10 ng | 466 | $3.35 \times 10^{10}$ | $2.15 \times 10^{6}$ | 4.90% | 0.006% | Amplicon | *E.coli* |
| ***TS-10*** | 5 ng | 466 | $1.67 \times 10^{10}$ | $8.49 \times 10^{5}$ | 7.10% | 0.005% | Amplicon | *E.coli* |
| ***TS-11*** | 1 ng | 466 | $3.35 \times 10^{9}$ | $8.67 \times 10^{4}$ | 21.90% | 0.003% | Amplicon | *E.coli* |
| ***TS-12*** | 500 pg | 466 | $1.67 \times 10^{9}$ | $1.31 \times 10^{5}$ | 7.60% | 0.008% | Amplicon | *E.coli* |
| ***IgG*** | 43 ng | 180 | 4.38E+11 | $3.24 \times 10^{6}$ | 17.61% | 0.001% | Shotgun | *Mus Musclus* |
| ***K27-1*** | 60 ng | 180 | 6.11E+11 | $1.87 \times 10^{6}$ | 9.40% | 0.001% | Shotgun | *Mus Musclus* |
| ***K27-2*** | 60 ng | 180 | 6.11E+11 | $1.49 \times 10^{6}$ | 5.65% | 0.001% | Shotgun | *Mus Musclus* |
| ***Ace*** | 723 ng | 550 | $2.42 \times 10^{12}$ | $3.63 \times 10^{8}$ | 6.10% | 0.015% | Shotgun | *A.longum* |
| ***pX*** | 60 ng | 180 | $6.11 \times 10^{11}$ | $9.06 \times 10^{6}$ | 4.60% | 0.001% | Shotgun | *Homo sapien* |

## Table 2: Trace library Generation & Sequence results Solexa

| Solexa Libraries | Input (ng) | DNA Library (total molecules by UT-dPCR)*/ul | Average number of clusters generated per tile | Total number of reads | % mapping to Human ref* |
|---|---|---|---|---|---|
| Plasma DNASample 1 | 3.2 | 1.07E+11 | 115998 | 11599833 | 51.46 |
| Plasma DNA Sample 2 | 3.6 | 7.88E+10 | 114548 | 11454876 | 52.66 |
| Plasma DNA Sample 3 | 2.7 | 7.17E+10 | 118516 | 11851612 | 56.05 |
| Plasma DNA Sample 4 | 2.6 | 6.03E+10 | 150414 | 15041417 | 49.67 |
| Plasma DNA Sample 5 | 5.6 | 7.17E+10 | 119104 | 11910483 | 56.13 |
| Plasma DNA Sample 6 | 2.4 | 7.23E+10 | 120974 | 12097478 | 55.39 |
| Whole Blood Genomic DNA Sample | 2.1 | 6.30E+10 | 151201 | 15120171 | 50.52 |

**Table 3: Trace library Sequence results 454 *Flx***

| ***Sample ID*** | ***Organism*** | ***Library Type*** | ***input (ng)*** | ***Proportion of library sequenced*** | ***Raw bases (Mbp)*** | **Number of reads** | ***Average read length (bp)*** | **% mapping to template/ assembling*** |
|---|---|---|---|---|---|---|---|---|
| ***TS-5*** | *E. coli* | Shotgun | 1.0 ng | 1.0 | 6.0 | 28,730 | 210.5 | 69.9% |
| ***TS-6*** | *E. coli* | Shotgun | 0.5 ng | 1.0 | 5.7 | 26,812 | 212.5 | 69.9% |
| ***TS-11*** | *E. coli* | Amplicon | 1.0 ng | 0.5 | 10.9 | 43,217 | 252.5 | 99.2% |
| ***TS-12*** | *E. coli* | Amplicon | 0.5 ng | 0.5 | 14.0 | 55,206 | 253.6 | 99.0% |
| ***IgG*** | *Mus Musclus* | Shotgun | 43 ng | 1.0 | 3.8 | 27,712 | 139.8 | 10.13% |
| ***K27-1*** | *Mus Musclus* | Shotgun | 60 ng | 1.0 | 4.1 | 27,701 | 147.7 | 35.34% |
| ***K27-2*** | *Mus Musclus* | Shotgun | 60 ng | 1.0 | 7.1 | 42,829 | 166.0 | 20.96% |
| ***pX*** | *Homo Sapien* | Shotgun | 60 ng | 0.21 | 42 | 244,010 | 172.6 | 64.6% |
| ***Ace*** | *A. longum* | Shotgun | 723 ng | 0.005 | 67 | 278,181 | 240.9 | 98.8%* |